\documentclass[a4paper,aps,prb,twocolumn,amsmath,amssymb,floatfix,preprintnumbers,footinbib,eqsecnum,groupedaddress]{revtex4}

\usepackage{amsmath}
\usepackage[pdftex]{graphicx}
\usepackage{dcolumn}
\usepackage{bm}
\usepackage{dsfont}
\usepackage{color}
\usepackage{pdfpages}

\newcounter{defcounter}
\usepackage{amsthm}

\begin{document}
\title{Dynamical Coulomb blockade of tunnel junctions driven by alternating voltages}
\author{Hermann Grabert}
\email{grabert@uni-freiburg.de}
\affiliation{Physikalisches Institut, Universit\"at Freiburg, Hermann-Herder-Stra{\ss}e 3, 79104 Freiburg, Germany}

\affiliation{Freiburg Institute for Advanced Studies (FRIAS), Albertstr.\ 19, 79104 Freiburg, Germany}

\begin{abstract}
The theory of the dynamical Coulomb blockade is extended to tunneling elements driven by a time-dependent voltage. It is shown that for standard set-ups where an external voltage is applied to a tunnel junction via an impedance, time-dependent driving entails an excitation of the modes of the electromagnetic environment by the applied voltage. Previous approaches for ac driven circuits need to be extended to account for the driven bath modes. A unitary transformation involving also the variables of the electromagnetic environment is introduced which allows us to split off the time dependence from the Hamiltonian in the absence of tunneling. This greatly simplifies perturbation-theoretical calculations based on treating the tunneling Hamiltonian as a perturbation. In particular, the average current flowing in the leads of the tunnel junction is studied. Explicit results are given for the case of an applied voltage with a constant dc part and a sinusoidal ac part. The connection with standard dynamical Coulomb blockade theory for constant applied voltage is established. It is shown that an alternating voltage source reveals significant additional effects caused by the electromagnetic environment. The hallmark of the dynamical Coulomb blockade in ac driven devices is a suppression of higher harmonics of the current by the electromagnetic environment. The theory presented basically applies to all tunneling devices driven by alternating voltages.  
\end{abstract}

\date{September 30, 2015;\
Published in:\ Phys.\ Rev.\ B {\bf 92}, 245433 (2015)}


\maketitle
\section{Introduction}\label{sec:one}

The dynamical Coulomb blockade (DCB) was one of the first phenomena revealing the quantum nature of electrical impedances. The theory of the DCB\cite{Devoret_1990,Girvin_1990,Ingold_1992}, also called $P(E)$ theory, explains the experimentally observed suppression of the low-voltage conductance of tunneling elements\cite{Delsing_1989,Geerligs_1989,Cleland_1992} as an effect of photon exchange between the tunneling element and its electromagnetic environment. It has found numerous applications to problems as diverse as atomic-sized contacts\cite{Agrait_2003}, artificial atoms\cite{Brandes_2005}, Josephson and Majorana qubits\cite{Martinis_2009,Rainis_2012}, current sources\cite{Pekola_2013} and the quantum measurement of work\cite{Pekola_2013b}. 
While standard $P(E)$ theory\cite{Devoret_1990,Girvin_1990,Ingold_1992} considers circuits with constant voltage bias, more recent  work\cite{Altland_2010,Safi_2010,Safi_2011,Galperin_2012,Souquet_2014,Gaury_2014} has addressed  nanostructures driven by alternating voltages inspired by nanoelectronic
experiments in the GHz range and above\cite{Shibata_2012,Gasse_2013,Parlavecchio_2015}. 

In this paper we extend $P(E)$ theory to tunnel junctions driven by alternating voltages. The modes of the environmental impedance, which play a crucial role for the conventional DCB effect, also cause significant effects in ac driven devices. Since an external voltage is applied via the environmental impedance and not directly to the junction electrodes, the voltage drives the modes of the environment. Driven bath modes can strongly influences the dynamics of a system.\cite{Grabert_2015} For an ac driven tunnel junction this implies additional DCB effects investigated in the sequel. 

The paper is organized as follows. In Sec.~\ref{sec:two} we recall the Hamiltonian for standard DCB theory. The model Hamiltonian is based on a representation of the environmental impedance as a collection of $LC$ circuits. We also introduce the relevant current operators of the circuit. In the presence of a time-dependent voltage the tunneling current and the current flowing in the leads need to be distinguished. Sec.~\ref{sec:three} presents the perturbation theory in the tunneling Hamiltonian, and we derive a general expression for the current flowing through the environmental impedance valid up to second order. We also give the unperturbed equations of motion of the circuit variables in the absence of tunneling. Results on the time evolution of the mean values provide the connection between the microscopic parameters in the Hamiltonian and phenomenological circuit theory. 

Due to the external driving, the unperturbed Hamiltonian is time-dependent which complicates the evaluation of the perturbation-theoretical results. In particular, an applied time-dependent voltage also drives the modes of the electromagnetic environment so that standard techniques for ac driven circuits fail. In Sec.~\ref{sec:four} we introduce a unitary transformation involving also the variables of the environmental modes which allows us to split off the time dependence arising from the external voltage. 

In Sec.~\ref{sec:five} we then analyze the perturbation-theoretical result of Sec.~\ref{sec:three} for the average current flowing in the leads. The relevant averages over unperturbed quantities factorize into averages over the quasiparticles in the junction leads and averages over the degrees of freedom of the environmental Hamiltonian. The quasiparticle averages are shown to reduce to those known from standard $P(E)$ theory. The averages over the environmental degrees of freedom can be reduced to averages over operators depending only on the phase operator which is conjugate to the junction charge. Employing the unitary transformation introduced previously and using the Gaussian nature of phase fluctuations in an undriven circuit in the absence of tunneling, all quantities can be expressed in terms of the phase-phase correlation function familiar from standard DCB theory.

In Sec.~\ref{sec:six} we study the average current in the presence of an applied voltage which is the sum of a constant dc part and a sinusoidal ac part. We determine the effective voltage across the tunnel junction which differs from the applied voltage because of a renormalization of the ac amplitude and a phase shift caused by the electromagnetic environment. The average current is also periodic and can be written as a Fourier series. We obtain a general expression for the Fourier coefficients. For the case of a constant applied voltage the results of standard $P(E)$ theory are recovered. The time-averaged current in the presence of sinusoidal driving is derived and related to the dc current for constant driving. The higher harmonics of the current are also determined and shown to be influenced by the environmental impedance implying a novel DCB effect in the ac range.  

Finally in Sec.~\ref{sec:seven} we present our conclusions and discuss the experimental observability of the predictions made in the previous sections. It is shown that for a tunnel junction driven through an $LC$ resonator the ac Coulomb blockade effect leads to a strong suppression of higher harmonics of the average current. This effect should  be readily observable with state-of-the-art setups.

\section{Model and Hamiltonian}\label{sec:two}
\begin{figure}
\includegraphics[width=0.38\textwidth]{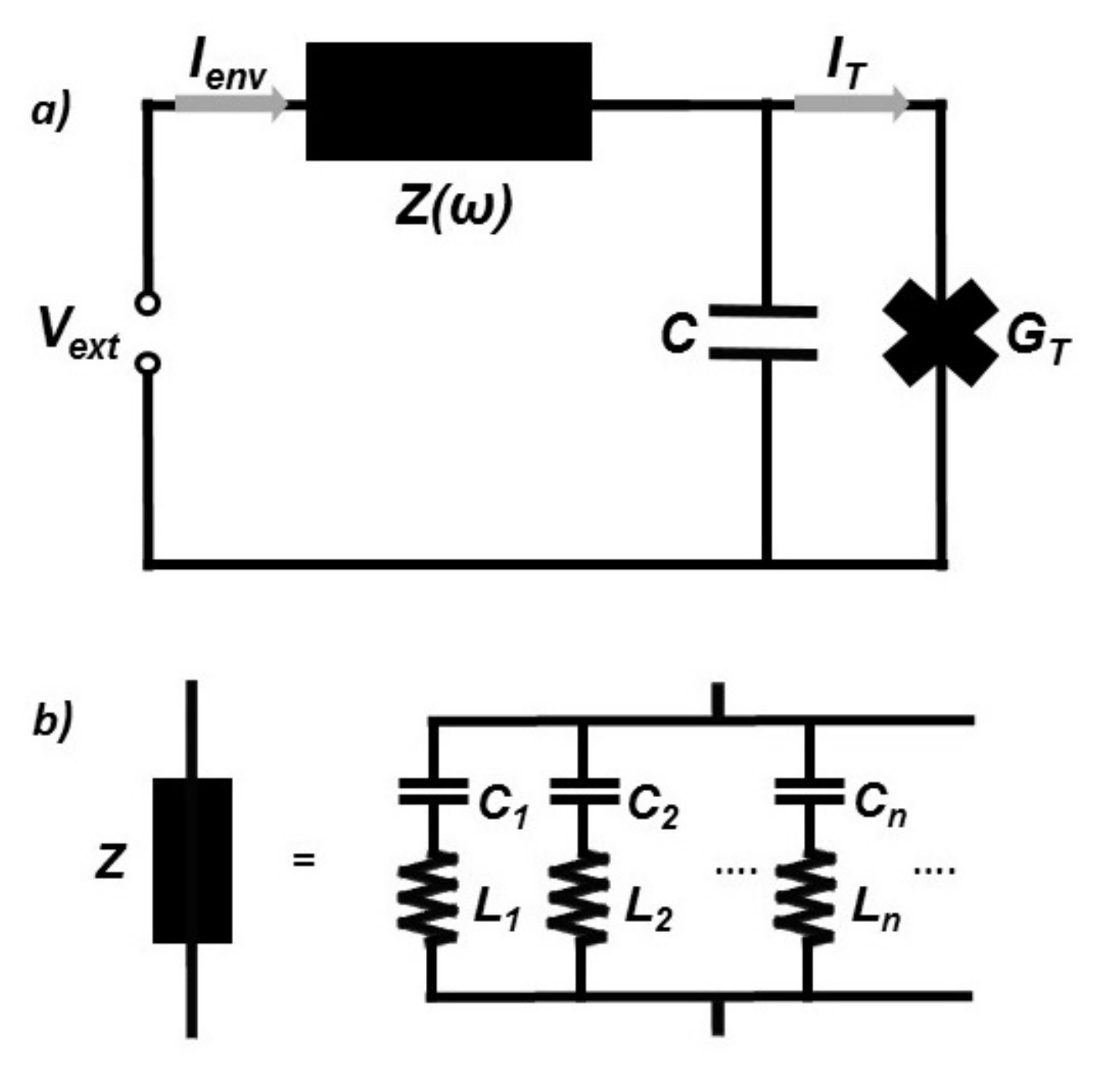}
\caption{\label{fig1} Circuit diagram of a voltage biased tunneling element.  a) General set-up showing a tunnel junction with capacitance $C$ and tunneling conductance $G_T$ coupled to a voltage source $V_{ext}$ via an external impedance $Z(\omega)$. The current $I_{env}$ flowing through the environmental impedance and the tunneling current $I_T$ are also indicated.  b) Representation of the impedance $Z(\omega)$ as a collection of $LC$ circuits.}
\end{figure}

We consider a tunnel junction with capacitance $C$ and tunneling conductance $G_T$  driven by a voltage source $V_{ext}(t)$ via an environmental impedance $Z(\omega)$. The circuit diagram of the system is displayed in Fig.\ref{fig1}. 
For a constant applied voltage  $V_{ext}(t)=V_{dc}$ this model coincides with the standard model for the DCB in tunnel junctions\cite{Devoret_1990,Girvin_1990,Ingold_1992}.

\subsection{Hamiltonian}
The Hamiltonian of the circuit may be written as
\begin{equation}\label{Htot}
H=H_{el} +H_T + H_{env}
\end{equation}
where
\begin{equation}
H_{el}=\sum_{k,\sigma}\epsilon_{k\sigma}a_{k\sigma}^{\dag}a_{k\sigma}^{}+
      \sum_{q,\sigma}\epsilon_{q\sigma}a_{q\sigma}^{\dag}a_{q\sigma}^{}
\end{equation}
describes the conduction electrons of the leads on either side of the tunnel junction.
Here $a_{k\sigma}^{}$ is the annihilation operator of an electron state with energy $\epsilon_{k\sigma}$ where $k$ denotes the longitudinal wave number and $\sigma$ denotes the transversal and spin quantum numbers. Likewise,  $a_{q\sigma}^{}$ is the annihilation operator of an electron state with energy $\epsilon_{q\sigma}$  in the other electrode, where $q$ denotes the longitudinal wave number. The quantum number  $\sigma$ is conserved during tunneling transitions described by the tunneling Hamiltonian
\begin{equation}\label{HT}
 H_T=\Theta \, e^{-i\varphi} + \Theta^{\dagger}e^{i\varphi}
\end{equation}
with the quasiparticle tunnel operator
\begin{equation}\label{tunop}
\Theta = \sum_{k,q,\sigma}t_{kq \sigma}a_{k\sigma}^{\dag}a_{q\sigma}^{}
\end{equation}
where the $t_{kq \sigma}$ are tunneling amplitudes. Furthermore, we have introduced the phase operator $\varphi$  conjugate to the junction charge $Q$. These operators obey the canonical
commutation relation\cite{Devoret_1990} $
\left[ \varphi, Q\right] = ie$.
Accordingly, the charge shift operator ${\rm e}^{-i\varphi}$ transfers one electronic charge across the junction. 
The last term in Eq.~(\ref{Htot}) describes the junction capacitance $C$ and the environmental impedance $Z(\omega)$ and may be written in the form\cite{Ingold_1992}
\begin{eqnarray}\label{Henv}
H_{env}(t) &=&\frac{Q^2}{2C} +\sum_n \Big\{\frac{Q_n^2}{2C_n} \\ \nonumber
&& +\frac{1}{2L_n} \left(\frac{\hbar}{e}\right)^2\left[\varphi-\varphi_n-\varphi_{ext}(t)\right]^2  \Big\}
\end{eqnarray}
where the first term is the charging energy of the tunnel junction with capacitance $C$ and
the second term represents the impedance $Z(\omega)$ in terms of a Caldeira-Leggett model\cite{Caldeira_1981,Caldeira_1983} as a collection of $LC$ circuits. These environmental modes represent a thermal bath causing dissipation and fluctuations and lead to the DCB effect. The charge operators $Q_n$ are conjugate to the phase operators $\varphi_n$ with the commutators
$\left[\varphi_n, Q_n \right] = ie$. The phase $\varphi_{ext}(t)$ describes the external driving and is defined by
\begin{equation}\label{phiext}
\dot \varphi_{ext}(t) = \frac{e}{\hbar} V_{ext}(t)
\end{equation}
where $V_{ext}(t)$ is the voltage applied to the circuit in Fig.\ref{fig1}. Hence the Hamiltonian $H_{env}$ depends on time $t$. The connection between the microscopic parameters in the Hamiltonian (\ref{Henv}) and the impedance $Z(\omega)$ of the circuit will become clear from the following.

\subsection{Current operators}
We first consider the electric currents flowing in the circuit. Using the Hamiltonian (\ref{Htot}), we find for the time rate of change of the junction charge
\begin{equation}
\dot Q = \frac{i}{\hbar}\left[ H,Q\right] = \frac{i}{\hbar}\left[ H_{env},Q\right] + \frac{i}{\hbar}\left[ H_T,Q\right]
\end{equation}
Inserting the explicit form of $H$, one obtains 
\begin{equation}\label{current}
\dot Q=I_{env}-I_T
\end{equation}
where\cite{Goppert_2000}
\begin{equation}\label{Ienv}
I_{env}=-\frac{\hbar}{e}\sum_n \frac{\varphi-\varphi_n-\varphi_{ext}}{L_n} 
\end{equation}
is the current flowing through the environmental impedance $Z(\omega)$ which on average charges the junction capacitance in case of an applied voltage, while
\begin{equation}\label{Itun2}
I_T= i\frac{e}{\hbar}\left[\Theta\, e^{-i\varphi}- \Theta^{\dagger}\, e^{i\varphi}\right]
\end{equation}
is the tunneling current which on average discharges the capacitance (see Fig.~\ref{fig1}).
For constant voltage bias $V_{dc}$ the average currents $\langle I_{env}\rangle$ and  $\langle I_T\rangle$ coincide. However, for time-dependent driving the two currents differ due to displacement currents flowing in the circuit and only $I_{env}$  is experimentally observable. Note that $I_{env}$ has an explicit time dependence via $\varphi_{ext}(t)$.

\section{Perturbation theory in the tunneling Hamiltonian}\label{sec:three}
$P(E)$ theory treats tunneling elements with weak tunneling conductance $G_T$. Hence we may consider the tunneling Hamiltonian $H_T$ as a small perturbation. 

\subsection{Expansion of time evolution operator}
To evaluate the time evolution for weak tunneling we write the total Hamiltonian (\ref{Htot}) as
\begin{equation}
H=H_0+H_T
\end{equation}
where
\begin{equation}
H_0=H_{el}+H_{env}
\end{equation}
is the Hamiltonian for $G_T=0$. Quite generally, a quantum mechanical operator $A$  reads in the Heisenberg representation
\begin{equation}\label{AHeis}
A(t)=U^{\dagger}(t,0)\, A\, U(t,0)
\end{equation}
where $t_0=0$ is the initial time when driving starts and
\begin{equation}
U(t,t^{\prime}) = \mathcal{T} \exp\left\{ -\frac{i}{\hbar} \int_{t^\prime}^t ds \, H(s) \right\}  
\end{equation}
is the time evolution operator with the time ordering operator $\mathcal{T}$. Expanding $U(t,0)$ up to second order in $H_T$ we have
\begin{eqnarray}\label{pert}
&&U(t,0)\doteq U_0(t,0)-\frac{i}{\hbar}\int_{0}^t ds\,U_0(t,s)H_TU_0(s,0)\\ \nonumber
&&-\frac{1}{\hbar^2}\int_{0}^t ds_1\int_{0}^{s_1} ds_2\,
U_0(t,s_1)H_TU_0(s_1,s_2)H_TU_0(s_2,0)
\end{eqnarray}
where the symbol $\doteq$ means equal apart from terms of higher order in $H_T$ and
\begin{equation}\label{U0}
U_0(t,t^{\prime}) = \mathcal{T} \exp\left\{ -\frac{i}{\hbar} \int_{t^\prime}^t ds \, H_0(s) \right\}  
\end{equation}
is the time evolution operator for $G_T=0$. 

Introducing the interaction representation of operators
\begin{equation}
\check A(t)=U_0^{\dagger}(t,0)\, A\, U_0(t,0)
\end{equation}
and inserting the result (\ref{pert}) into Eq.~(\ref{AHeis}), we find for the Heisenberg operator $A(t)$ in second order in $H_T$ the result
\begin{eqnarray}\label{AHeis2}
&&A(t) \doteq \check A(t)   
+\frac{i}{\hbar} \int_{0}^t ds\,   \left[\check H_T(s),  \check A (t)\right] \\ \nonumber 
&& \qquad -\frac{1}{\hbar^2}\int_{0}^t ds_1 \int_{0}^{s_1} ds_2\,   \left[\check H_T(s_2),  \left[\check H_T(s_1),  \check A (t)\right] \right]
\end{eqnarray}
This general expression can now be used to obtain an explicit result for the current $I_{env}$ in second order in $H_T$. Inserting the form (\ref{HT}) of the tunneling Hamiltonian, we find
\begin{eqnarray}\label{IenvHeis}\nonumber
&&I_{env}(t) \doteq \check I_{env}(t)   \\ \nonumber
&&\qquad
+\frac{i}{\hbar} \int_{t_0}^t ds\,   \left[\check\Theta(s)\, e^{-i\check\varphi(s)}+ \hbox{H.c.}\, ,  \check I_{env}  (t)\right] \\ \nonumber
&&\qquad -\frac{1}{\hbar^2}\int_{t_0}^t ds_1 \int_{t_0}^{s_1} ds_2\,   \Big[\check\Theta(s_2)\, e^{-i\check\varphi(s_2)}+ \hbox{H.c.}\, , \\ 
&&\qquad\qquad \Big[\check\Theta(s_1)\, e^{-i\check\varphi(s_1)}+ \hbox{H.c.}\, ,  \check I_{env}  (t)\Big] \Big]
\end{eqnarray}
Here the operators $\check \Theta(t)$, $\check\varphi(t)$ and $\check I_{env}(t)$ are in the interaction picture. Hence, their time dependence is governed by the Hamiltonian $H_0=H_{el}+H_{env}$ in the absence of tunneling.

Since the circuit described by $H_{env}$ and the electrodes described by $H_{el}$ are decoupled for vanishing tunneling, the time evolution operator (\ref{U0}) factorize according to
\begin{equation}\label{Ufac}
U_0(t,t^{\prime})=U_{env}(t,t^{\prime}) \, e^{-\frac{i}{\hbar}H_{el}(t-t^{\prime})}
\end{equation}
into the time evolution operator 
\begin{equation}\label{Uenv}
U_{env}(t,t^{\prime}) = \mathcal{T} \exp\left\{ -\frac{i}{\hbar} \int_{t^\prime}^t ds \, H_{env}(s) \right\}  
\end{equation}
of the electrodynamic environment and the time evolution operator of the quasiparticles which takes a simple form since $H_{el}$ is independent of time also in the presence of driving. 

\subsection{Heisenberg equations of motion in the absence of tunneling}

The evaluation of the perturbative results requires an explicit solution of the unperturbed problem.
The Heisenberg equations of motion 
\begin{equation}
\dot {\check A} =\frac{i}{\hbar}\left[ H_0,\check A\right]
\end{equation}
of the circuit variables in the interaction picture are found to read
\begin{equation}\label{eomphi}
\dot {\check \varphi} =\frac{e}{\hbar}\,\frac{\check Q}{C}, 
\end{equation}
\begin{equation}\label{eomQ}
\dot {\check Q} =   -\frac{\hbar}{e}\sum_n \frac{\check\varphi-\check\varphi_n-\varphi_{ext}}{L_n} 
\end{equation}
and
\begin{equation} \label{eomphin}
\dot {\check \varphi}_n = \frac{e}{\hbar}\,\frac{\check Q_n}{C_n}, 
\end{equation}
\begin{equation}\label{eomQn}
\dot  {\check Q}_n =\frac{\hbar}{e}\, \frac{\check \varphi - \check \varphi_n-\varphi_{ext}}{L_n}
\end{equation}
Introducing the frequencies
\begin{equation}
\omega_n=\frac{1}{\sqrt{L_nC_n}}
\end{equation}
of the $LC$ circuits of the Caldeira-Leggett model and
combining Eqs.~(\ref{eomphin}) and (\ref{eomQn}) we readily obtain
\begin{equation}\label{ddotphin}
\ddot {\check \varphi}_n  + \omega_n^2 \check \varphi_n= \omega_n^2 \left(\check \varphi-\varphi_{ext}\right)
\end{equation}
This evolution equation can easily be solved to yield
\begin{eqnarray}
&&\check \varphi_n(t) =\varphi_n(0) \cos\left(\omega_n t\right) + \frac{e}{\hbar}\,\frac{Q_n(0)}{\omega_n C_n}\sin\left(\omega_n t\right) \qquad\\ \nonumber
&&\qquad +\omega_n \int_0^t ds \sin\left[\omega_n(t-s)\right]\left[\check\varphi(s)-\varphi_{ext}(s)\right]
\end{eqnarray}
When this is combined with Eqs.~(\ref{eomphi}) and (\ref{eomQ}) we obtain after a partial integration
\begin{eqnarray}\label{langevin}
&&C\ddot{\check\varphi}(t)  +\int_0^t ds \,Y(t-s)\dot{\check \varphi}(s)\\ \nonumber
&&\qquad = \frac{e}{\hbar} i_N(t) + 
\int_0^t ds \,Y(t-s)\dot\varphi_{ext}(s)
\end{eqnarray}
where we have introduced the temporal response function\cite{Ingold_1992}
\begin{equation}\label{Yoft}
Y(t) = \sum_n \frac{1}{L_n} \cos(\omega_nt)
\end{equation}\noindent
of the electromagnetic environment.
Furthermore, the evolution equation (\ref{langevin}) includes a quantum noise current
\begin{eqnarray} \nonumber\label{xi}
i_N(t)  &=& \sum_n \frac{1}{L_n}\bigg\{\frac{\hbar}{e}\left[ \varphi_n(0)-\varphi(0)+\varphi_{ext}(0)\right]\cos(\omega_nt) \\
&&\quad +\frac{Q_n(0)}{\omega_nC_n}\sin(\omega_nt)\bigg\}
\end{eqnarray}
depending on the initial state of the environmental modes. 

For later purposes we also note that in the absence of tunneling the current (\ref{Ienv}) may be written as
\begin{equation}\label{checkIenv}
\check I_{env}(t) = \dot{\check Q}(t)=\frac{\hbar}{e}C\ddot{\check \varphi}(t)
\end{equation}
as is readily seen by combining Eqs. (\ref{Ienv}), (\ref{eomphi}) and (\ref{eomQ}).

\subsection{Time evolution of mean values in the absence of tunneling}
Let us assume that the system has reached the equilibrium state in the absence of driving at time $t=0$ and is then driven out of equilibrium for times $t>0$. Then $\varphi_{ext}(t)=0$ for $t\le 0$, and from the Hamiltonian (\ref{Henv}) with $\varphi_{ext}(t)=0$ we see that in the initial equilibrium state
\begin{equation}
\langle Q_n(0) \rangle = 0
\end{equation}
and
\begin{equation}
\langle \varphi_n(0)-\varphi(0) \rangle = 0
\end{equation}
This implies in view of Eq.~(\ref{xi}) and $\varphi_{ext}(0)=0$
\begin{equation}
\langle i_N(t) \rangle = 0
\end{equation}
From Eq.~(\ref{langevin}) we thus obtain for the average motion
\begin{equation}\label{average}
C\langle \ddot{\check \varphi}(t) \rangle +\int_0^t\!\! ds \, Y(t-s)\langle\dot{\check\varphi}(s)\rangle =
\int_0^t\!\!  ds \, Y(t-s)\, \dot\varphi_{ext}(s)
\end{equation}
Using now Eqs.~(\ref{phiext}) and (\ref{eomphi}) this may be written as
\begin{equation}\label{pheno}
\langle \dot{\check Q}(t) \rangle =\int_0^t ds \,Y(t-s)\left[V_{ext}(s) -\frac{\langle \check Q(s)\rangle }{C} \right]
\end{equation} 
The evolution equation (\ref{pheno}) describes the circuit of Fig.\ref{fig1} for $G_T=0$ phenomenologically.  Since in the absence of tunneling $\langle \dot{ \check Q}\rangle$ is the current in the circuit and $V_{ext}-\langle\check Q\rangle/C$ the voltage across the environmental impedance, Eq.~(\ref{pheno}) is just the current-voltage relation of the environmental impedance. Accordingly, the admittance
\begin{equation}\label{Yofom}
Y(\omega)=1/Z(\omega)=\int_0^{\infty}dt\, Y(t)\, e^{i\omega t}
\end{equation}
 is the Fourier transform of $Y(t)$. Hence, Eq.~(\ref{Yoft}) relates the parameters of the Caldeira-Leggett model with the phenomenological theory of the circuit.

\section{Unitary transformation}\label{sec:four}

 In the presence of an applied voltage the Hamiltonian $H_{env}$ is time-dependent which complicates perturbation-theoretical calculations. For constant voltage the time-dependence caused by $\varphi_{ext}(t)$ can be shifted by a unitary transformation\cite{Ingold_1992} to a phase factor of the tunneling Hamiltonian $H_T$.  A related unitary transformation of the form $\mathcal{U}(t)=e^{-iQ\Phi(t) }$ still exists\cite{Safi_2010,Safi_2011} if a time-dependent voltage is applied directly to the two electrodes on either side of the tunnel junction. However, when the external voltage is applied via an environmental impedance $Z(\omega)$, as it is the case for the circuit in Fig.\ref{fig1}, the voltage also drives the environmental modes as is explicitly seen from Eq.~(\ref{ddotphin}). Accordingly, the time-dependence can only be split off by means of a unitary transformation involving the environmental modes as well.

To find the proper transformation, we make the ansatz
\begin{equation}
\check \varphi(t)= \bar\varphi(t) +\tilde \varphi(t)
\end{equation}
\begin{equation}
\check Q(t)= \bar Q(t)+\tilde Q(t)
\end{equation}
where
\begin{equation}
\bar Q(t) =\frac{\hbar}{e}C\dot{\bar\varphi}(t) 
\end{equation}
and
\begin{equation}
\check \varphi_n(t)= \bar\varphi_n(t) +\tilde \varphi_n(t)
\end{equation}
\begin{equation}
\check Q_n(t)= \bar Q_n(t)+\tilde Q_n(t)
\end{equation}
where
\begin{equation}
\bar Q_n(t)= \frac{\hbar}{e}C_n\dot{\bar\varphi}_n (t) 
\end{equation}

Inserting this into the equations of motion (\ref{eomphi}) -- (\ref{eomQn}), one finds that the operators $\tilde\varphi$, $\tilde Q$, $\tilde\varphi_n$, and $\tilde Q_n$ satisfy the equations of motion for vanishing driving provided $\bar\varphi(t)$ and $\bar\varphi_n(t)$ obey
\begin{equation}\label{eomphires}
\ddot{\bar\varphi}+ \sum_n\frac{\bar \varphi - \bar \varphi_n}{L_nC}
=\sum_n\frac{\varphi_{ext}}{L_nC}
\end{equation}
and
\begin{equation}\label{eomphinres}
\ddot{\bar\varphi}_n+\omega_n^2\left(\bar\varphi_n -\bar\varphi \right)=-\omega_n^2\varphi_{ext}
\end{equation}
The particular solution of Eq.~(\ref{eomphinres}) with initial conditions $\bar\varphi(0)=0$ and  $\dot{\bar\varphi}(0)=0$ reads
\begin{equation}\nonumber
\bar\varphi_n(t)= \omega_n\int_0^t ds\, \sin\left[\omega_n(t-s)\right]\left[\bar\varphi(s)-\varphi_{ext}(s)\right]
\end{equation}
When this is inserted into Eq.~(\ref{eomphires}) one finds after a partial integration using $\bar\varphi(0)=0$ and the definition (\ref{Yoft}) of $Y(t)$ the equation of motion for $\bar\varphi(t)$ in the form
\begin{equation}\label{eomphires2}
C\ddot{\bar\varphi}(t) + \int_0^t ds\, Y(t-s)\left[\dot{\bar\varphi}(s)-\dot\varphi_{ext}(s)\right] =0
\end{equation}
which coincides with the equation of motion  (\ref{average}) for the average phase $\langle \check \varphi(t) \rangle$.

It is now convenient to introduce the unitary transformation 
\begin{eqnarray}\label{Lambda}
\Lambda(t)&=&\exp\left\{ \frac{i}{e}\left[\bar\varphi(t)Q +\sum_n\bar \varphi_n(t)Q_n  \right]\right\} \\ \nonumber && \times
\exp\left\{-\frac{i\hbar}{e^2}\left[C\dot{\bar\varphi}(t)\varphi+\sum_n C_n\dot{\bar\varphi}_n(t)\varphi_n\right]\right\}
\end{eqnarray}
This leads to the following transformation of the circuit operators
\begin{eqnarray}\label{trafphi}
\hat \varphi&=& \Lambda(t) \varphi \, \Lambda^{\dagger}(t) = \varphi +\bar\varphi(t)\\
\nonumber
\hat Q&=& \Lambda(t) Q \, \Lambda^{\dagger}(t) = Q + \frac{\hbar}{e}C\dot{\bar\varphi}(t)\\
\nonumber
\hat \varphi_n&=& \Lambda(t) \varphi_n \, \Lambda^{\dagger}(t) = \varphi_n +\bar\varphi_n(t)\\
\nonumber
\hat Q_n&=& \Lambda(t) Q_n \, \Lambda^{\dagger}(t) = Q_n + \frac{\hbar}{e}C_n\dot{\bar\varphi}_n(t)
\end{eqnarray}
When we apply this time-dependent change of variables, the Hamiltonian must be transformed as
\begin{equation}\label{trafHenv}
\hat H_{env}= \Lambda(t) H_{env} \,\Lambda^{\dagger}(t) + i \hbar \frac{\partial \Lambda(t)}{\partial t}\Lambda^{\dagger}(t)
\end{equation}
From Eq.~(\ref{Lambda}) one obtains
\begin{eqnarray}\nonumber\label{LambdaLambda}
&& i \hbar \frac{\partial \Lambda(t)}{\partial t}\Lambda^{\dagger}(t) 
=\left(\frac{\hbar}{e}\right)^2C\ddot{\bar \varphi}\left(\varphi+\bar\varphi\right)-\frac{\hbar}{e}\dot{\bar\varphi}\, Q \\
&&+\sum_n\left[\left(\frac{\hbar}{e}\right)^2C_n\ddot{ \bar\varphi}_n\left(\varphi_n+\bar\varphi_n\right) - \frac{\hbar}{e}\dot{\bar\varphi}_n Q_n \right]
\end{eqnarray}
We now insert Eq.~(\ref{LambdaLambda}) and the Hamiltonian (\ref{Henv}) into Eq.~(\ref{trafHenv}). One then finds by virtue of the transformation laws (\ref{trafphi})  and by exploiting Eqs.~(\ref{eomphires}) and (\ref{eomphinres})  satisfied by $\bar\varphi$ and $\bar\varphi_n$ after some algebra
\begin{equation}
\hat H_{env} = H^0_{env} + G(t)
\end{equation}
where
\begin{equation}\label{Henv0}
H^0_{env} =\frac{Q^2}{2C}
+\sum_n \bigg[\frac{Q_n^2}{2C_n} 
+\frac{1}{2L_n}\left(\frac{\hbar}{e}\right)^2 \left(\varphi-\varphi_n\right)^2  \bigg] 
\end{equation}
is the Hamiltonian of the electromagnetic environment in the absence of driving and
\begin{eqnarray}\nonumber
G(t)&=&\left(\frac{\hbar}{e}\right)^2 \left[C\bar\varphi\ddot {\bar\varphi}+ \frac{1}{2}C\dot{\bar\varphi}^2\right]
+\sum_n \left(\frac{\hbar}{e}\right)^2 \bigg[C_n\bar\varphi_n\ddot {\bar\varphi}_n \\ 
&&
+\frac{1}{2}C_n\dot{\bar\varphi}_n^2
+\frac{1}{2L_n} \left(\bar\varphi-\bar\varphi_n-\varphi_{ext}\right)^2\bigg] 
\end{eqnarray}
Hence, apart from the time-dependent function $G$ the Hamiltonian $H_{env}$ has been transformed to the time-independent Hamiltonian (\ref{Henv0}) of an undriven circuit.

Regarding the time evolution operator (\ref{Uenv}) of the driven circuit, we can now employ the unitary transformation (\ref{Lambda}) to obtain
\begin{equation}\label{URLC1}
U_{env}(t,t^{\prime}) = \Lambda^{\dagger}(t) \, e^{-\frac{i}{\hbar}H_{env}^0(t-t^{\prime})}\Lambda(t^{\prime}) \,e^{i\delta(t,t^{\prime})}
\end{equation}
where
\begin{equation}
\delta(t,t^{\prime})= -(1/\hbar)\int_{t^{\prime}}^t ds\, G(s)
\end{equation}
is a phase factor.

\section{Average current}\label{sec:five}
After the preparations in the previous sections the average current $\langle I_{env}(t)\rangle$ can be evaluated. When we determine the average current from Eq.~(\ref{IenvHeis}), only terms involving a product of the quasiparticle tunnel operator $\Theta$ and its adjoint $\Theta^{\dagger}$ give a non-vanishing contribution, since all other terms do not conserve the number of quasiparticles in each electrode. We thus obtain for the average current
\begin{eqnarray}\label{Iav}
&&\langle I_{env}(t) \rangle \doteq \langle \check I_{env}(t) \rangle  -\frac{1}{\hbar^2}\int_{0}^t ds_1 \int_{0}^{s_1} ds_2\, 
 \\ \nonumber
&&\Big\{\left\langle \left[ \check\Theta(s_2)\, e^{-i\check\varphi(s_2)}\, ,
\left[\check\Theta^{\dagger}(s_1)\, e^{i\check\varphi(s_1)}, \check I_{env}(t) \right]\right]\right\rangle+\hbox{c.c.} \Big\}
\end{eqnarray}
These averages factorize into a quasiparticle average and an average of the electromagnetic environment. In view of the factorization (\ref{Ufac}) of the unperturbed time evolution operator, the interaction representation of the operators in Eq.~(\ref{Iav}) reads
\begin{equation}
\check \Theta(t) = e^{\frac{i}{\hbar}H_{el}t} \,\Theta\, e^{-\frac{i}{\hbar}H_{el}t}
\end{equation} 
\begin{equation}
\check \Theta^{\dagger}(t) = e^{\frac{i}{\hbar}H_{el}t} \,\Theta^{\dagger}\, e^{-\frac{i}{\hbar}H_{el}t}
\end{equation} 
\begin{equation}
\check \varphi(t) = U^{\dagger}_{env}(t,0) \, \varphi \, U_{env}(t,0)
\end{equation} 
and
\begin{equation}
\check I_{env}(t) = U^{\dagger}_{env}(t,0) \, I_{env} \, U_{env}(t,0)
\end{equation} 
Using now the representation (\ref{URLC1}) of $U_{env}(t,t^{\prime})$, we obtain by virtue of Eq.~(\ref{trafphi})
\begin{equation}\label{phicheck}
\check \varphi(t) = e^{\frac{i}{\hbar}H_{env}^0t} \, \varphi \, e^{-\frac{i}{\hbar}H_{env}^0t} + \bar\varphi(t)
\end{equation}
where we have taken into account that at time $t_0=0$ the transformation (\ref{Lambda}) is trivial, i.e. $\Lambda(0)=1$, since driving is absent for $t\leq 0$. Introducing the Heisenberg operators of the undriven electromagnetic environment
\begin{equation}
\tilde A(t) =  e^{\frac{i}{\hbar}H_{env}^0t} \, A\, e^{-\frac{i}{\hbar}H_{env}^0t}
\end{equation}
Eq.~(\ref{phicheck}) simply reads
\begin{equation}\label{phicheck2}
\check \varphi(t) = \tilde \varphi(t) +\bar \varphi(t)
\end{equation}
Likewise we find for the interaction representation of the current operator (\ref{Ienv})
\begin{equation}\label{Ienvcheck}
\check I_{env}(t) =  \tilde I_{env}(t)  + \bar I_{env}(t)
\end{equation}
where
\begin{equation}\label{tildeIenv}
\tilde I_{env}(t) = -\frac{\hbar}{e}\sum_n\frac{\tilde\varphi(t)-\tilde \varphi_n(t)}{L_n} =\dot{\tilde Q}(t) =\frac{\hbar}{e}C\ddot{\tilde\varphi}(t)
\end{equation}
is the current operator in the absence of driving and tunneling and
\begin{eqnarray}\label{barIenv}\nonumber
\bar I_{env}(t)&=&  -\frac{\hbar}{e}\sum_n\frac{\bar\varphi(t)-\bar \varphi_n(t)-\varphi_{ext}(t)}{L_n}\\ && =\dot{\bar Q}(t)  =\frac{\hbar}{e}C\ddot{\bar\varphi}(t)
\end{eqnarray}
is the average displacement current flowing in a driven circuit in the absence of tunneling. The relations of the components $\tilde I_{env}$ and $\bar I_{env}$ of $\check I_{env}$ to those of $\check Q$ and $\check \varphi$ follow in analogy to Eq.~(\ref{checkIenv}).

\subsection{Quasiparticle averages}

The quasiparticle averages contributing to the average current (\ref{Iav}) are easily evaluated. 
Introducing the function
\begin{equation}\label{alpha}
\alpha(t-s)= \left\langle\check\Theta(t)\,  \check\Theta^{\dagger}(s) \right\rangle 
\end{equation}
and inserting the explicit form of the tunnel operator (\ref{tunop}) one obtains
\begin{eqnarray}\label{ThetaTheta}
\alpha(t)=\sum_{k,q,\sigma}\vert t_{kq \sigma}\vert^2 \left\langle \check a_{k\sigma}^{\dag}(t) \check a_{q\sigma}^{}(t) a_{q\sigma}^{\dag} a_{k\sigma}^{} \right\rangle
\end{eqnarray}
Here, the average is over the quasiparticle equilibrium state $\rho_{\beta}^{el}=(1/Z_{el})\exp(-\beta H_{el})$ of the electrodes where $\beta=1/k_BT$ is the inverse temperature. The evaluation of the quantity (\ref{ThetaTheta}) for a tunnel junction is familiar\cite{Ingold_1992,Grabert_1994}.  In the wide band limit  the tunneling amplitudes squared $\vert t_{kq\sigma}\vert^2$ may be replaced by an average $\overline{t}^2$. Inserting then the quasiparticle Greens functions one obtains
\begin{equation}
\alpha(t)= \overline{t}^2\sum_{k,q,\sigma}\frac{e^{\frac{i}{\hbar}\left(\epsilon_{k\sigma}-\epsilon_{q\sigma}\right)t}}{\left(1+e^{\beta\epsilon_{k\sigma}}\right)\left(1+e^{-\beta\epsilon_{q\sigma}}\right)}
\end{equation}
The index $\sigma$ labels a large number ${\cal N} \gg 1$ of channels, and the sums over $k$ and $q$ may be replaced by integrals over energy when we introduce the densities of states $\rho$ and $\rho^{\prime}$ of the two electrodes. One then finds 
\begin{equation}\label{ThetaTheta3}
\alpha(t) =  \overline{t}^2{\cal N}\rho\rho^{\prime}\int_{-\infty}^{\infty} dE\, \frac{E\, e^{-\frac{i}{\hbar}E t}}{1-e^{-\beta E}} 
\end{equation}
Now, the tunneling conductance $G_T$ is given by\cite{Ingold_1992}
\begin{equation}
G_T = \frac{2\pi e^2}{\hbar}\overline{t}^2{\cal N}\rho\rho^{\prime}
\end{equation}
in terms of which
the result (\ref{ThetaTheta3}) takes the form 
\begin{equation}\label{ThetaTheta4}
\alpha(t) = \frac{1}{2\pi}\frac{\hbar}{e^2}G_T\int_{-\infty}^{\infty} dE\, \frac{E\, e^{-\frac{i}{\hbar}E t} }{1-e^{-\beta E}} 
\end{equation}
The other quasiparticle averages in Eq.~(\ref{Iav}) can be evaluated in the same way.
One obtains 
\begin{equation}
\left\langle\check\Theta^{\dagger}(t) \check\Theta(s) \right\rangle = \alpha(t-s)
\end{equation}
and 
\begin{equation}
\left\langle\check\Theta(s) \check\Theta^{\dagger}(t) \right\rangle 
= \left\langle\check\Theta^{\dagger}(s) \check\Theta(t) \right\rangle = \alpha^*(t-s)
\end{equation}
Hence, the arising quasiparticle averages coincide with those known from standard $P(E)$ theory.

\subsection{Phase averages}
The phase averages required for a circuit driven by a time-dependent voltage are more involved  since $I_{env}$ is of order zero in the quasiparticle tunneling operator $\Theta$ while for standard $P(E)$ theory\cite{Ingold_1992} it suffices to compute the average of  $I_T$ which is explicitly of first order in $\Theta$.
Writing the quasiparticle averages in terms of the function $\alpha(t)$,  the average current (\ref{Iav}) reads
\begin{eqnarray}\label{Ienv2}\nonumber
\langle I_{env}(t) \rangle &\doteq& \langle \check I_{env}(t) \rangle  -\frac{1}{\hbar^2}\int_{t_0}^t ds_1 \int_{t_0}^{s_1} ds_2\, \bigg\{ \alpha(s_1-s_2)
 \\ 
&& \times\Big(\left\langle \left[ \check I_{env}(t), e^{i\check\varphi(s_1)}\right]  e^{-i\check\varphi(s_2)}\right\rangle \\ \nonumber
&&+ \left\langle \left[ \check I_{env}(t), e^{-i\check\varphi(s_1)}\right]  e^{i\check\varphi(s_2)}\right\rangle \Big)+ \hbox{c.c.}\bigg\}
\end{eqnarray}
where we have rewritten the result for an arbitrary initial time $t_0$ when driving starts.
What remains to be evaluated are averages of phase operators in the interaction representation.

We are interested in the current when the driving has been acting for a long time. Then, we may let the initial time tend to $-\infty$ and find
\begin{eqnarray}\label{Ienvav3}
\langle I_{env}(t) \rangle &\doteq& \langle \check I_{env}(t) \rangle  -\frac{1}{\hbar^2}\int_0^{\infty} du \int_0^{\infty} dv\, \bigg\{ \alpha(v)
 \\ \nonumber
&& \times\Big(\left\langle \left[ \check I_{env}(t), e^{i\check\varphi(t-u)}\right]  e^{-i\check\varphi(t-u-v)}\right\rangle \\ \nonumber
&&+ \left\langle \left[ \check I_{env}(t), e^{-i\check\varphi(t-u)}\right]  e^{i\check\varphi(t-u-v)}\right\rangle \Big)+ \hbox{c.c.}\bigg\}
\end{eqnarray}
We now take advantage of the unitary transformation (\ref{Lambda}) which leads to the representations (\ref{phicheck2}) and (\ref{Ienvcheck}) of the phase operators in the interaction representation. Inserting these formulas we obtain
\begin{eqnarray}\label{Ienvav4}
\langle I_{env}(t) \rangle &\doteq&  \bar I_{env}(t)   -\frac{1}{\hbar^2}\int_0^{\infty} du \int_0^{\infty} dv\, \bigg\{ \alpha(v)
 \\ \nonumber
&& \times\left\langle \left[ \tilde I_{env}(t), e^{i\tilde\varphi(t-u)}\right]  e^{-i\tilde\varphi(t-u-v)}\right\rangle \\ \nonumber
&&\qquad \times\left( e^{i[\bar\varphi(t-u)-\bar\varphi(t-u-v)]}-\hbox{c.c.}\right)+ \hbox{c.c.}\bigg\}
\end{eqnarray}
where all expectation values are now reduced to averages to be determined for the undriven electromagnetic environment described by the Hamiltonian $H_{env}^0$ introduced in Eq.~(\ref{Henv0}). We also have taken advantage of the fact that the Hamiltonian $H_{env}^0$ is invariant under the parity transformation $\Pi$ with the properties
\begin{eqnarray}
\Pi&=&\Pi^{\dagger},\quad \Pi^2=1\\
\Pi\, \varphi\, \Pi &=& -\varphi,\qquad \Pi\, \varphi_n\, \Pi = -\varphi_n\\
\Pi\, Q\, \Pi &=& -Q,\qquad \Pi\, Q_n\, \Pi = -Q_n
\end{eqnarray}
This implies $ \Pi\, H_{env}^0\, \Pi =H_{env}^0$ and the relations
\begin{eqnarray}
&&\left\langle  e^{-i\tilde\varphi(t)}\,   e^{i\tilde\varphi(t-s)}   \right\rangle  =\left\langle  e^{i\tilde\varphi(t)}\,   e^{-i\tilde\varphi(t-s)}   \right\rangle  
\end{eqnarray}
and
\begin{eqnarray}
&&\left\langle \left[ \tilde I_{env}(t), e^{i\tilde\varphi(t-u)}\right]  e^{-i\tilde\varphi(t-u-v)}\right\rangle \\ \nonumber
&&\quad =-\left\langle \left[ \tilde I_{env}(t), e^{-i\tilde\varphi(t-u)}\right]  e^{i\tilde\varphi(t-u-v)}\right\rangle 
\end{eqnarray}
This latter relation has been used to simplify the result (\ref{Ienvav4}).

\subsection{Phase correlation function}

The evaluation of phase averages is familiar from the standard theory of the DCB\cite{Devoret_1990,Girvin_1990,Ingold_1992}. Since the process of phase fluctuations of the undriven environment is stationary and Gaussian, one has
\begin{equation}\label{Gauss}
\left\langle  e^{-i\tilde\varphi(t)}\,   e^{i\tilde\varphi(t-s)}   \right\rangle =
e^{ \left\langle \left[\tilde\varphi(s) -\tilde\varphi(0)\right]  \tilde\varphi(0)   \right\rangle} =e^{J(s)}
\end{equation}
where 
\begin{equation}\label{Joft}
J(t)=\left\langle  \left[ \tilde\varphi(t)- \tilde\varphi(0) \right] \tilde\varphi(0)  \right\rangle
\end{equation}
is the phase-phase correlation function of the undriven circuit which takes the form\cite{Ingold_1992}
\begin{eqnarray}\label{Joft2}
J(t) &=& 2 \int_0^{\infty} \frac{d\omega}{\omega}\frac{\hbox{ Re}\left[Z_t(\omega)\right]}{R_K}\\ \nonumber
&&\quad\times \left\{
\coth\left(\frac{1}{2}\beta\hbar\omega\right)\left [\cos(\omega t) - 1\right] - i \sin(\omega t)\right\}
\end{eqnarray}
where $R_K=h/e^2$ is the resistance quantum and 
\begin{equation}\label{Zt}
Z_t(\omega)=\frac{1}{Y(\omega)-i\omega C}
\end{equation}
is the total impedance of the circuit consisting of the environmental impedance and the junction capacitance.

To determine the average current of a circuit driven by a time-dependent voltage one needs to evaluate the phase average
\begin{equation}
\left\langle \left[ \tilde I_{env}(t), e^{i\tilde\varphi(u)}\right]  e^{-i\tilde\varphi(v)}\right\rangle  =\frac{\hbar}{e}C\left\langle \left[ \ddot{\tilde\varphi} (t), e^{i\tilde\varphi(u)}\right]  e^{-i\tilde\varphi(v)}\right\rangle
\end{equation}
where we have used Eq.~(\ref{tildeIenv}).
Due to the Gaussian statistics of phase fluctuations in the absence of tunneling we have
\begin{eqnarray}
&&\left\langle  \tilde\varphi(t) \, e^{i\tilde\varphi(u)}\,  e^{-i\tilde\varphi(v)}\right\rangle \\ \nonumber
&&\quad = i \left[\left\langle \tilde\varphi(t) \tilde\varphi(u)\right\rangle 
- \left\langle \tilde \varphi(t) \tilde\varphi(v) \right\rangle\right]\left\langle  e^{i\tilde\varphi(u)}\,  e^{-i\tilde\varphi(v)}\right\rangle\\ \nonumber
&&\quad = i \left[J(t-u)-J(t-v)\right] e^{J(u-v)}
\end{eqnarray}
and
\begin{eqnarray}
&&\left\langle e^{i\tilde\varphi(u)}\,  \tilde\varphi(t) \,  e^{-i\tilde\varphi(v)}\right\rangle \\ \nonumber
&&\quad = i \left[\left\langle  \tilde\varphi(u) \tilde\varphi(t)\right\rangle 
- \left\langle \tilde \varphi(t) \tilde\varphi(v) \right\rangle\right]\left\langle  e^{i\tilde\varphi(u)}\,  e^{-i\tilde\varphi(v)}\right\rangle\\ \nonumber
&&\quad = i \left[J(u-t)-J(t-v)\right] e^{J(u-v)}
\end{eqnarray}
which gives
\begin{eqnarray}\label{phasecomm}\nonumber
&&\left\langle \left[ \tilde\varphi(t) , e^{i\tilde\varphi(u)}\right]  e^{-i\tilde\varphi(v)}\right\rangle\\ 
&&\quad  = i \left[J(t-u)-J(u-t)\right] e^{J(u-v)} \\ 
&&\quad = -2J^{\prime\prime}(t-u) \, e^{J(u-v)} \nonumber
\end{eqnarray}
where we have used the symmetry 
\begin{equation}\label{Joftsym}
J(-t)=J^*(t)
\end{equation}
obeyed by the phase-phase correlation function (\ref{Joft2}) and have introduced the imaginary part $J^{\prime\prime}(t)$ of $J(t)$ with the second-order time derivative
\begin{equation}\label{Joft3}
\ddot J^{\prime\prime}(t) = \frac{e^2}{\pi \hbar} \int_0^{\infty} d\omega\,\omega\hbox{ Re}\left[Z_t(\omega)\right]  \sin(\omega t)
\end{equation}
By virtue of Eq.~(\ref{phasecomm}) we now obtain
\begin{equation}
\left\langle \left[ \tilde I_{env}(t), e^{i\tilde\varphi(u)}\right]  e^{-i\tilde\varphi(v)}\right\rangle =-\frac{2\hbar C}{e}\ddot J^{\prime\prime}(t-u) \, e^{J(u-v)}
\end{equation}
Inserting this result into Eq.~(\ref{Ienvav4}), we find for the average current
\begin{eqnarray}\label{Ienvav5}
&&\langle I_{env}(t) \rangle \doteq  \bar I_{env}(t)   +\frac{2C}{\hbar e}\int_0^{\infty} \!\!\!\! du \int_0^{\infty}\!\!\!\! dv\, \bigg\{ \alpha(v)\ddot J^{\prime\prime}(u)\, e^{J(v)}
 \nonumber \\ 
&&\qquad\times\left( e^{i[\bar\varphi(t-u)-\bar\varphi(t-u-v)]}-\hbox{c.c.}\right)+ \hbox{c.c.}\bigg\}
\end{eqnarray}
Here $\bar I_{env}(t)$ is the average displacement current flowing in the absence of tunneling and the additional terms of second order in $H_T$ contain the correlator $\alpha(t)$ from the average over the quasiparticles, the correlator $J(t)$ from the phase averages in the absence of driving, and exponential factors taking account of the applied voltage (\ref{Vext}) which has been split off from the averages by means of the unitary transformation (\ref{Lambda}).

\section{Average tunneling current for periodic driving}\label{sec:six}

We focus now on the important case of a circuit driven by the voltage
\begin{equation}\label{Vext}
V_{ext}(t)=V_{dc} + V_{ac}\cos\left(\Omega t\right)
\end{equation}
comprising a dc voltage $V_{dc}$ and an ac voltage $V_{ac}$ of frequency $\Omega$. From Eq.~(\ref{phiext}) we find for the phase $\varphi_{ext}(t)$ associated with the voltage drive (\ref{Vext}) apart from an arbitrary additive constant
\begin{eqnarray}\label{drive}
\varphi_{ext}(t) &=& \frac{e}{\hbar}\left[V_{dc}t + \frac{V_{ac}}{\Omega}\sin\left(\Omega t\right) \right]\\ \nonumber
&=& \frac{e}{\hbar}\left[V_{dc}t -\hbox{Re} \left\{\frac{V_{ac}}{i\Omega}\, e^{-i\Omega t}\right\} \right]
\end{eqnarray}
Combining this with the evolution equation (\ref{eomphires2}) for  $\bar \varphi(t)$  we obtain 
\begin{equation}
\bar \varphi(t) =\frac{e}{\hbar}\left[V_{dc} t
-{\rm Re} \left\{\frac{ Y(\Omega)}{
 Y(\Omega)- i\Omega C}\frac{V_{ac}}{i\Omega}\,e^{-i\Omega t}\right\}\right]
\end{equation}
when the periodic driving (\ref{drive}) has been acting for a sufficiently long time so that a steady oscillatory state has been reached. We split $\bar\varphi(t)$ into
\begin{equation}\label{phisplit}
\bar\varphi(t) = \frac{e}{\hbar}V_{dc}t + \bar\varphi_{ac}(t)
\end{equation}
with the ac component
\begin{equation}\label{phiac}
\bar \varphi_{ac}(t) =-\frac{e}{\hbar}{\rm Re} \left\{\frac{  Y(\Omega)}{
Y(\Omega)- i\Omega C}\frac{V_{ac}}{i\Omega}\,e^{-i\Omega t}\right\}
\end{equation}
Since $\bar\varphi(t)$ coincides with the average phase $\langle\check\varphi(t)\rangle$ in the absence of tunneling, we find by virtue of Eq.~(\ref{eomphi}) for the average charge $\langle \check Q(t)\rangle$ the result 
\begin{equation}\label{Qav}
\langle \check Q(t) \rangle = CV_{dc} +{\rm Re} \left\{\frac{  CY(\Omega)}{
Y(\Omega)- i\Omega C}V_{ac}\,e^{-i\Omega t}\right\}
\end{equation}
where $Y(\omega)=1/Z(\omega)$ is the admittance of the electrodynamic environment. In the absence of tunneling the junction charge $\langle \check Q(t) \rangle$ oscillates with an amplitude proportional to $V_{ac}$ about
the time-averaged charge $CV_{dc}$. Introducing the polar decomposition 
\begin{equation}\label{polarde}
 \frac{Y(\Omega)}{Y(\Omega)-i\Omega C} =\Xi(\Omega) \, e^{i\eta(\Omega)}
\end{equation}\noindent
of the admittance ratio into modulus $\Xi$ and phase $\eta$, Eq.~(\ref{Qav}) simply reads 
\begin{equation}\label{Qav2}
\langle \check Q(t) \rangle = C [V_{dc} + V_{\Omega}\cos(\Omega t -\eta)]
\end{equation}
which shows that the ac voltage across the junction has a renormalized  amplitude
\begin{equation}\label{VOm}
V_{\Omega} = \Xi\, V_{ac}
\end{equation}
and is shifted by the phase $\eta$ due to the electromagnetic environment.

Making use of the polar decomposition 
(\ref{polarde}) and the effective ac voltage (\ref{VOm}) across the junction, the phase (\ref{phiac}) takes the form
\begin{equation}\label{phiac2}
\bar \varphi_{ac}(t) =\frac{eV_{\Omega}}{\hbar\Omega}\sin(\Omega t -\eta)
\end{equation}
Accordingly, we find
\begin{equation}\label{phidif}
\bar\varphi(t)-\bar\varphi(t-s)=\frac{e}{\hbar}V_{dc}s +a\sin(\Omega t -\eta) - a\sin[\Omega (t-s) -\eta]
\end{equation}
where we have introduced the amplitude
\begin{equation}
a=\frac{eV_{\Omega}}{\hbar\Omega} 
\end{equation}
The phase difference (\ref{phidif}) can now be inserted into Eq.~(\ref{Ienvav5}) to give for the average current  $\langle I_{env}(t) \rangle $ the result 
\begin{eqnarray}\label{Ienvav6}
&&\!\!\!\! \langle I_{env}(t) \rangle \doteq  \bar I_{env}(t)   +\frac{2C}{\hbar e}\int_0^{\infty}\!\!\! du \int_0^{\infty}\!\!\! dv\, \bigg\{ \alpha(v)\ddot J^{\prime\prime}(u)\, e^{J(v)}
 \nonumber \\ \nonumber
&&\quad \times\left( e^{\frac{i}{\hbar}eV_{dc}v}e^{ia\sin[\Omega(t-u)-\eta]}e^{-ia\sin[\Omega(t-u-v)-\eta]}-\hbox{c.c.}\right)\\ 
&&\qquad + \hbox{c.c.}\bigg\}
\end{eqnarray}
This result has the same periodicity in time as the ac driving voltage.

\subsection{Fourier components of average current}

To evaluate the expression (\ref{Ienvav6}) for the average current further, we  employ the Jacobi-Anger expansion\cite{Abramowitz} of exponentials of trigonometric functions.
One has
\begin{equation}\label{Bessel}
e^{ia\sin(\Omega t)}=\sum_{k=-\infty}^{\infty}
J_k(a)\, e^{ik\Omega t}
\end{equation}
where $J_k(z)$ is the Bessel function of the first kind of order $k$. 
For real $a$ and integer $k$ the Bessel functions are real and obey
\begin{equation}
J_{-k}(a) = (-1)^k J_k(a) = J_k(-a)
\end{equation}
Using the series representation (\ref{Bessel}), one obtains for the average current (\ref{Ienvav6})
\begin{eqnarray}\label{Ienvav7}\nonumber
&&\langle I_{env}(t) \rangle \doteq  \bar I_{env}(t)   +\frac{2C}{\hbar e}\sum_{k,l=-\infty}^{\infty}\int_0^{\infty} du \int_0^{\infty} dv\, \\ 
&&\quad\bigg\{ \alpha(v)\ddot J^{\prime\prime}(u)\, e^{J(v)} J_k(a)J_l(a)
\\ \nonumber
&&\quad \times\left( e^{\frac{i}{\hbar}eV_{dc}v}e^{i(k-l)[\Omega(t-u)-\eta]}e^{il\Omega v}-\hbox{c.c.}\right) \, + \hbox{c.c.}\bigg\}
\end{eqnarray}
This can be written as a Fourier series
\begin{equation}\label{FS}
\langle I_{env}(t)\rangle =\sum_{n=-\infty}^{\infty} I_n\, e^{-in(\Omega t -\eta)}
\end{equation}
The Fourier coefficients may be decomposed into
\begin{equation}
I_n=\bar I_n + \hat I_n
\end{equation}
where the $\bar I_n$ describe the current $\bar I_{env}(t)$ which coincides with the displacement current in the absence of tunneling, while the coefficients $\hat I_n$ arise from tunneling. From Eq.~(\ref{Qav2}) we obtain for the displacement current $\bar I_{env}(t) = \langle \check I_{env}(t) \rangle= \langle \dot{\check Q}(t) \rangle$  the expression
\begin{equation}
\bar I_{env}(t) = -\Omega CV_{\Omega}\sin(\Omega t -\eta)
\end{equation}
from which we see that $\bar I_{env}$ gives only a contribution to the coefficients $I_{\pm 1}$ of the form
\begin{equation}\label{barI1}
\bar I_1 = -\bar I_{-1}=-\frac{i}{2}\Omega CV_{\Omega}
\end{equation}
The Fourier coefficients $\hat I_n$ due to tunneling transitions can be read off from Eq.~(\ref{Ienvav7})
with the result
\begin{eqnarray}\label{hatIn}
\hat I_n&\doteq&  \frac{2C}{\hbar e}\sum_{k=-\infty}^{\infty}\int_0^{\infty} du\, \ddot J^{\prime\prime}(u)\, e^{in\Omega u}\int_0^{\infty} dv\,\\ \nonumber
&& \bigg\{ \alpha(v)\, e^{J(v)}\Big( J_{k-n}(a)J_{k}(a)\, e^{\frac{i}{\hbar}(eV_{dc}+k\hbar\Omega) v} \\ \nonumber
&&\quad - J_{k+n}(a)J_{k}(a)\, e^{-\frac{i}{\hbar}(eV_{dc}+k\hbar\Omega) v} \Big) \\ \nonumber
&&+  \alpha^*(v)\, e^{J^*(v)}\Big( J_{k+n}(a)J_{k}(a)\, e^{-\frac{i}{\hbar}(eV_{dc}+k\hbar\Omega )v}\\ \nonumber
&&\quad   - J_{k-n}(a)J_{k}(a)\, e^{\frac{i}{\hbar}(eV_{dc}+k\hbar\Omega) v} \Big)\, \bigg\}
\end{eqnarray}
To evaluate this further we make use of the representation (\ref{Joft3}) of $\ddot J^{\prime\prime}(t)$ which implies
\begin{eqnarray}\label{ddotJoft}
&&\int_0^{\infty} du \,\ddot J^{\prime\prime}(u) \, e^{in\Omega u}\\  \nonumber
&&\quad=
\frac{e^2}{\pi \hbar} \int_0^{\infty} du   \int_0^{\infty} d\omega\,\omega\hbox{ Re}\left[Z_t(\omega)\right]  \sin(\omega u) \, e^{in\Omega u}\\  \nonumber
&&\quad =
\frac{e^2}{2\pi \hbar} \int_0^{\infty} du   \int_{-\infty}^{\infty} d\omega\,\omega\, Z_t(\omega) \sin(\omega u) \, e^{in\Omega u}
\end{eqnarray}
where we have used  the symmetry $Z_t^*(\omega)=Z_t(-\omega)$ of the total impedance (\ref{Zt}) to obtain the last line. With the help of
\begin{eqnarray}\nonumber
&&\!\!\!\! \int_0^{\infty}\!\!\! du\,   \sin(\omega u) \, e^{in\Omega u}= \frac{1}{2i}\left[\pi \delta(\omega+n\Omega) -\pi \delta(\omega-n\Omega)\right]\\ 
&&\qquad\qquad +\frac{1}{2}P \left[\frac{1}{\omega+n\Omega}+\frac{1}{\omega-n\Omega}\right] 
\end{eqnarray}
where $P$ denotes the Cauchy principal value, the result (\ref{ddotJoft}) simplifies to read
\begin{eqnarray}\label{factor}\nonumber
&&\int_0^{\infty} du \,\ddot J^{\prime\prime}(u) \, e^{in\Omega u}=
\frac{e^2}{2 \hbar C}+i\frac{ n\Omega e^2}{2 \hbar } Z_t(n\Omega) \\
&&\qquad\qquad  =\frac{e^2}{2 \hbar C}\frac{Y(n\Omega)}{Y(n\Omega)-in\Omega C}
\end{eqnarray}
where we have inserted the explicit form (\ref{Zt}) of the total impedance to obtain the last expression. The result (\ref{factor}) can now be used to write the Fourier coefficients (\ref{hatIn}) in the form
\begin{eqnarray}\label{hatIn3}
&&\hat I_n \doteq \frac{e}{\hbar^2} \frac{Y(n\Omega)}{Y(n\Omega)-in\Omega C}\sum_{k=-\infty}^{\infty}\int_0^{\infty}\! \! ds \\ \nonumber
&&\quad\Big\{ \alpha(s)\, e^{J(s)}\Big(J_{k-n}(a)J_{k}(a) 
\,e^{\frac{i}{\hbar}(eV_{dc}+k\hbar\Omega)s}  \\ \nonumber
&&\qquad-J_{k+n}(a)J_{k}(a) \,e^{-\frac{i}{\hbar}(eV_{dc}+k\hbar\Omega)s}\Big)
\\ \nonumber
&&\quad +\alpha^*(s)\, e^{J^*(s)}\Big(J_{k+n}(a)J_{k}(a) 
\,e^{-\frac{i}{\hbar}(eV_{dc}+k\hbar\Omega)s}  \\ \nonumber
&&\qquad-J_{k-n}(a)J_{k}(a) \,e^{\frac{i}{\hbar}(eV_{dc}+k\hbar\Omega)s}\Big)\Big\}
\end{eqnarray}
These coefficients obey
\begin{equation}
\hat I_{-n}^*=\hat I_n^{}
\end{equation}
which in conjunction with Eq.~(\ref{barI1}) ensures that the current (\ref{FS}) is real.

\subsection{Time-averaged current}
Let us first consider the case $V_{ac}=0$. Then $a=0$ and since $J_k(0)=0$ for $k\neq 0$ while $J_0(0)=1$, only the Fourier coefficient $I_0$ is nonvanishing and we obtain for the stationary current $I_0$ in the presence of a dc voltage only
\begin{equation}\label{Idc}
I_{dc}(V_{dc}) \doteq \frac{e}{\hbar^2} 
\int_{-\infty}^{\infty} ds \,\alpha(s)\, e^{J(s)}
\left(e^{\frac{i}{\hbar}eV_{dc}s}-\hbox{c.c.}\right)
\end{equation}
where we have made use of the symmetry~(\ref{Joftsym}) of $J(t)$ and the relation
\begin{equation}\label{alphasym}
\alpha^*(t)=\alpha(-t)
\end{equation}
following from Eq.~(\ref{ThetaTheta4}).
Inserting the representation (\ref{ThetaTheta4}) of $\alpha(t)$ into Eq.~(\ref{Idc}) we obtain
\begin{eqnarray}\label{Idc2}
&& I_{dc}(V_{dc}) \doteq 
\frac{G_T}{2\pi \hbar e}\int_{-\infty}^{\infty} dE\, \frac{E }{1-e^{-\beta E}} \\ \nonumber
&&\quad \times 
\int_{-\infty}^{\infty}\! \! ds \, e^{-\frac{i}{\hbar}E s}\, e^{J(s)}\Big(
e^{\frac{i}{\hbar}eV_{dc}s}  - e^{-\frac{i}{\hbar}eV_{dc}s}
\Big)
\end{eqnarray}
Introducing now the familiar $P(E)$ function\cite{Devoret_1990,Ingold_1992}
\begin{equation}\label{PofE}
P(E)=\frac{1}{2\pi\hbar}\int_{-\infty}^{\infty}dt\, e^{J(t)+\frac{i}{\hbar}Et}
\end{equation}
which gives the probability to exchange the energy $E$  with the environmental modes during a tunneling transition, the expression (\ref{Idc2}) takes the form
\begin{eqnarray}\label{Idc3}\nonumber
&&I_{dc}(V_{dc}) \doteq  \frac{G_T}{e}
 \int_{-\infty}^{\infty} dE\, \frac{E }{1-e^{-\beta E}}\\
&&\qquad
\times \left[P(eV_{dc}-E)-P(-eV_{dc}-E)\right]
\end{eqnarray}
which is the standard result of dynamical Coulomb blockade theory\cite{Devoret_1990,Girvin_1990,Ingold_1992}.

Let us now turn again to the case of an ac drive and consider the time-averaged current $I_0=\hat I_0$. From Eq.~(\ref{hatIn3})   we obtain the result
\begin{eqnarray}\label{Inaught}
I_0 &\doteq& \frac{e}{\hbar^2} 
\sum_{k=-\infty}^{\infty}\int_{-\infty}^{\infty} ds \,\alpha(s)\, e^{J(s)}\\ \nonumber
&&\times J_{k}^2(a) 
\left(e^{\frac{i}{\hbar}(eV_{dc}+k\hbar\Omega)s}-\hbox{c.c.}\right)
\end{eqnarray}
We can now combine Eqs.~(\ref{Idc}) and (\ref{Inaught}) to obtain
\begin{equation}\label{I02}
I_0= \sum_{k=-\infty}^{\infty} J_k^2\left(\frac{e\Xi V_{ac}}{\hbar\Omega}\right)
 I_{dc}\left(V_{dc}+k\hbar\Omega/e\right)
\end{equation}
This corresponds to the result of Tien-Gordon theory\cite{Tien_1963} for the photon assisted dc tunneling current but it includes the lead impedance causing a renormalization of the effective ac voltage by the factor $\Xi$. The result (\ref{I02}) shows that the dc current of the circuit driven by a dc and a sinusoidal ac voltage can be determined from the current-voltage characteristics of the device driven by a dc voltage only\cite{Safi_2010,Safi_2011,Parlavecchio_2015,Tucker_1985}.

\subsection{Higher harmonics of average current}
Let us now consider the alternating part of the average current for sinusoidal driving. The time-dependent part of $\langle I_{env}(t)\rangle$  is described by the Fourier coefficients (\ref{barI1}) and (\ref{hatIn3}) for $n\neq 0$. It is convenient to write the result (\ref{hatIn3}) for the Fourier coefficients $\hat I_n$ in the form
\begin{eqnarray}\label{hatIn4}
&&\hat I_n \doteq \frac{e}{2\hbar^2} \frac{Y(n\Omega)}{Y(n\Omega)-in\Omega C}\sum_{k=-\infty}^{\infty}\int_{-\infty}^{\infty}\! \! ds \\ \nonumber
&&\quad\Big\{ \alpha(s)\, e^{J(s)} \left[J_{k+n}(a) +J_{k-n}(a) \right]J_{k}(a)\\ \nonumber
&&\quad\qquad\times
\left(e^{\frac{i}{\hbar}(eV_{dc}+k\hbar\Omega)s}-\hbox{c.c.}\right)
\\ \nonumber
&&\quad + \operatorname{sign}(s)\, \alpha(s)\, e^{J(s)}\left[ J_{k-n}(a)-J_{k+n}(a) \right]J_{k}(a)\\ \nonumber
&&\quad\qquad\times
\left(e^{\frac{i}{\hbar}(eV_{dc}+k\hbar\Omega)s}+\hbox{c.c.}\right)\Big\}
\end{eqnarray}
where we have decomposed the integrand into real and imaginary parts by making use of the symmetries~(\ref{Joftsym}) and (\ref{alphasym}) of $J(t)$  and $\alpha(t)$, respectively. It can now readily be seen that the real part of the integral in Eq.~(\ref{hatIn4}) can be expressed in terms of the current $I_{dc}(V_{dc})$ in the presence of a dc voltage only. Using Eq.~(\ref{Idc}) we find
\begin{eqnarray}\nonumber\label{intre}
&&\frac{e}{2\hbar^2}\sum_{k=-\infty}^{\infty}\int_{-\infty}^{\infty}\! \! ds \, \alpha(s)\, e^{J(s)} \left[J_{k+n}(a) +J_{k-n}(a) \right]J_{k}(a)\\ 
&&\quad\qquad\times
\left(e^{\frac{i}{\hbar}(eV_{dc}+k\hbar\Omega)s}-\hbox{c.c.}\right)\\  \nonumber
&&= \frac{1}{2}\sum_{k=-\infty}^{\infty}\left[J_{k+n}(a) +J_{k-n}(a)\right]J_{k}(a)
I_{dc}(V_{dc}+k\hbar\Omega/e)
\end{eqnarray}
To express also the imaginary part of the integral in Eq.~(\ref{hatIn4}) in terms of the dc current we introduce
\begin{equation}\label{IKK}
I_{KK}(V_{dc}) = \frac{ie}{\hbar^2} 
\int_{-\infty}^{\infty}\!\!\! ds \,\operatorname{sign}(s)\,\alpha(s)\, e^{J(s)}
\left(e^{\frac{i}{\hbar}eV_{dc}s}+\hbox{c.c.}\right)
\end{equation}
in terms of which we have
\begin{eqnarray}\nonumber\label{intim}
&&\!\!\!\! \frac{e}{2\hbar^2} \sum_{k=-\infty}^{\infty}\int_{-\infty}^{\infty}\! \! ds \, \operatorname{sign}(s)\, \alpha(s)\, e^{J(s)}\left[ J_{k-n}(a)-J_{k+n}(a) \right]\\ 
&&\quad\qquad\times J_{k}(a)
\left(e^{\frac{i}{\hbar}(eV_{dc}+k\hbar\Omega)s}+\hbox{c.c.}\right)\qquad \\  \nonumber
&&\!\!=  \frac{i}{2} 
\sum_{k=-\infty}^{\infty} \left[J_{k+n}(a) -J_{k-n}(a) \right]J_{k}(a)
I_{KK}(V_{dc}+k\hbar\Omega/e)
\end{eqnarray}
$I_{KK}(V)$ is the Kramers-Kronig transform of $I_{dc}(V)$ and can be determined by the relation\cite{Tucker_1985}
\begin{equation}\label{IKK2}
I_{KK}(V)= P\int_{-\infty}^{\infty}\frac{dU}{\pi} \frac{I_{dc}(U)-G_TU}{U-V}
\end{equation}
To see this one inserts the representation (\ref{Idc}) of $I_{dc}(V)$ into Eq.~(\ref{IKK2}) and makes use of
\begin{equation}
P\int_{-\infty}^{\infty}\frac{dU}{\pi} \frac{e^{\frac{i}{\hbar}eUs}-\hbox{c.c.}}{U-V} =i \operatorname{sign}(s)\left(e^{\frac{i}{\hbar}eVs}+\hbox{c.c.}\right)
\end{equation}
to recover the expression (\ref{IKK}). Furthermore, we note that the Fourier coefficients are not affected by a constant shift of $I_{KK}(V)$. The definition (\ref{IKK2}) of the Kramers-Kronig transformed current $I_{KK}(V)$ removes the asymptotic behavior of $I_{dc}(V)$ and leads to a well-behaved principal value integral.

With the help of the relations (\ref{intre}) and (\ref{intim}) one obtains from Eq.~(\ref{hatIn4}) 
\begin{eqnarray}\label{hatIn5}
&&\hat I_n \doteq \frac{1}{2} \frac{Y(n\Omega)}{Y(n\Omega)-in\Omega C}\sum_{k=-\infty}^{\infty} J_{k}(a)\\ \nonumber
&&\qquad\times\Big\{ \left[J_{k+n}(a) +J_{k-n}(a)\right]
I_{dc}(V_{dc}+k\hbar\Omega/e)\\ \nonumber
&&\qquad\ +i \left[J_{k+n}(a) -J_{k-n}(a) \right]
I_{KK}(V_{dc}+k\hbar\Omega/e)\Big\}
\end{eqnarray}
which combines with Eq.~(\ref{barI1}) to give for
the Fourier coefficients $I_n$ for $n>0$ the result 
\begin{eqnarray}\label{Infin}
&&I_n = \frac{1}{2} \frac{Y(n\Omega)}{Y(n\Omega)-in\Omega C}\sum_{k=-\infty}^{\infty} J_{k}(a)\\ \nonumber
&&\qquad\times\big\{ \left[J_{k+n}(a) +J_{k-n}(a)\right]
I_{dc}(V_{dc}+k\hbar\Omega/e)\\ \nonumber
&&\qquad\ +i \left[J_{k+n}(a) -J_{k-n}(a) \right]
I_{KK}(V_{dc}+k\hbar\Omega/e)\big\} \\ \nonumber
&&\qquad
-\frac{i}{2}\delta_{n,1} \Omega C \Xi V_{ac}
\end{eqnarray}
The last term present for $n=1$ only comes from the displacement current flowing already in the absence of tunneling. The coefficients for $n<0$ follow from $I_{-n}^{}=I_n^*$.

Equations (\ref{I02}) and (\ref{Infin}) determine the current $\langle I_{env}(t)\rangle $ caused by a dc and a sinusoidal ac voltage in terms of the current driven by a dc voltage only and the admittance of the electromagnetic environment.  The components (\ref{Infin}) of the alternating current reveal additional effects of the electromagnetic environment. Compared to previous work on ac voltages applied directly to the junction electrodes\cite{Safi_2010,Safi_2011,Parlavecchio_2015,Tucker_1985}, we find a suppression of the $n$th harmonic of the current by the factor $Y(n\Omega)/[Y(n\Omega)-in\Omega C]$.

\section{Conclusions}\label{sec:seven}

We have demonstrated that for ac driven tunneling elements there are additional effects of the electromagnetic environment not taken into account in previous work. For alternating voltages it is important to distinguish between the tunneling current flowing across the tunneling element and the experimentally observable current flowing in the leads of the tunnel junction. This distinction is only insignificant for devices driven by a constant voltage bias when no displacement currents flow in the circuit. 

An applied alternating voltage does not only affect the leads of the tunnel junction but also drives the modes of the electromagnetic environment. Therefore, the time dependence of the  Hamiltonian of the circuit arising from the applied voltage cannot be shifted to the tunneling Hamiltonian by a unitary transformation leaving the environmental modes unaffected. We have presented a unitary transformation of all circuit variables allowing us to reduce the unperturbed quantities to be evaluated in a perturbative expansion in the tunneling Hamiltonian to quantities calculable for a circuit in the absence of tunneling and driving. 

Regarding the average current driven by a constant dc and a sinusoidal ac voltage, the electromagnetic environment causes the following effects: The time-averaged current  (\ref{I02})
is essentially of the form of the photon assisted tunneling current of Tien-Gordon theory,\cite{Tien_1963} however, with an effective ac voltage $V_{\Omega}=\Xi V_{ac}$.
This is natural, since in the absence of tunneling the average voltage across the tunnel junction follows from Eq.~(\ref{Qav2}) as
\begin{equation}\label{VJ}
 V_J  =\frac{\langle \check Q(t) \rangle}{C}= V_{dc} +V_{\Omega}\cos(\Omega t -\eta)
\end{equation}
and thus has an ac amplitude modified by the electromagnetic environment. 
More pronounced are the modifications of the higher harmonics  (\ref{Infin}) of the current. Typically, these harmonics are suppressed since an alternating component of the tunneling current must partially be used to charge the junction capacitance thereby building up an alternating voltage which can drive the current in the lead circuit. Treating the average voltage (\ref{VJ}) across the tunnel junction in the absence of tunneling formally as an externally applied voltage, one can determine the average tunneling current   $\langle I_T \rangle$ perturbatively along the lines
of previous work\cite{Safi_2010,Safi_2011,Parlavecchio_2015,Tucker_1985}. Subsequently, circuit theory can be employed to decompose the tunneling current into a displacement current component and the current $\langle I_{env}\rangle $ flowing in the leads. This leads to results fully consistent with our findings. However, the approach presented here is not restricted to average quantities but treats tunneling elements driven by ac voltages on the level of current operators and can therefore likewise be applied to determine noise properties.

To illustrate the environmental effects let us briefly discuss two experimentally relevant cases. For a purely Ohmic environmental impedance, i.e., $Z(\omega)=R$, the ratio between the total impedance (\ref{Zt}) and the lead  impedance $Z(\omega)$  is given by
\begin{equation}\nonumber
\frac{Z_t(\omega)}{Z(\omega)}=\frac{Y(\omega)}{Y(\omega)-i\omega C}=\frac{1}{1-i\omega RC}
\end{equation}
The polar decomposition of this ratio gives for the modulus
\begin{equation}
\Xi(\omega) =\frac{1}{\sqrt{1+(\omega RC)^2}}
\end{equation}
and for the phase
\begin{equation}
\eta(\omega)=\arctan(\omega RC)
\end{equation}
Hence the modulus $\Xi$ is always smaller than 1. However, in particular for a typical lead resistance in the range of 50 $\Omega$, the suppression will mostly be moderate.

This changes if we consider a tunnel junction driven through a resonator as studied recently.\cite{Parlavecchio_2015,Altimiras_2014} 
\begin{figure}
\includegraphics[width=0.45\textwidth]{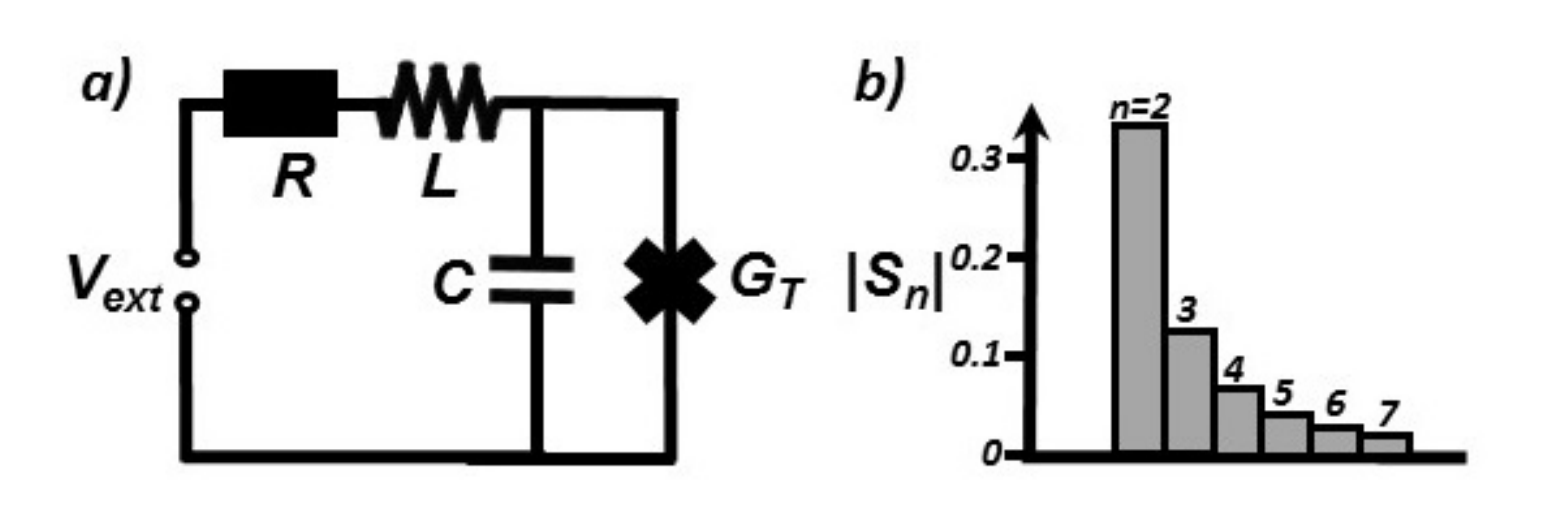}
\caption{\label{fig2} a) Circuit diagram of a tunnel junction driven through an $LC$ resonator with lead resistance $R$. b) Suppression factor $\vert S_n\vert$ of amplitude of higher harmonics of order $n$ for driving at the resonance frequency and a quality factor $Q_f=10$.}
\end{figure}
The circuit diagram of the set-up depicted in Fig.~\ref{fig2}a shows an environmental impedance
\begin{equation}
Z(\omega)=R-i\omega L
\end{equation}
with an Ohmic lead resistance $R$ and an inductance $L$. The $LC$ resonator of the circuit has a resonance frequency 
\begin{equation}
\omega_0=\frac{1}{\sqrt{LC}}
\end{equation}
and a characteristic impedance 
\begin{equation}
Z_c=\sqrt{\frac{L}{C}}
\end{equation}
implying a quality factor $Q_f=Z_c/R$ and accordingly a loss factor
\begin{equation}
\gamma=\frac{1}{Q_f}=\frac{R}{Z_c}
\end{equation}
For this circuit the ratio between the total impedance (\ref{Zt}) and the impedance $Z(\omega)$ of the leads is given by
\begin{equation}
\frac{Z_t(\omega)}{Z(\omega)}=\frac{1}{1-i\omega RC -\omega^2 LC}=\frac{\omega_0^2}{\omega_0^2-\omega^2-i\gamma\omega_0\omega}
\end{equation}
which implies a modulus
\begin{equation}\label{Xires}
\Xi(\omega)=\frac{\omega_0^2}{\sqrt{\left(\omega_0^2-\omega^2\right)^2+\left(\gamma\omega_0\omega\right)^2}} 
\end{equation}
and a phase
\begin{equation} \label{etares}
\eta(\omega)=\arctan\left(\frac{\gamma\omega_0\omega}{\omega_0^2-\omega^2}\right)
\end{equation}
where the values of $\arctan$ are to be chosen in the interval $[0,\pi)$.

The effects of the electromagnetic environment are most pronounced when the circuit is driven at the resonance frequency $\omega_0$ of the $LC$ resonator.  For a voltage of the form (\ref{Vext}) with an ac component of amplitude $V_{ac}$ and frequency $\Omega=\omega_0$, we obtain from Eqs.~(\ref{Xires}) and (\ref{etares}) for $\omega=\omega_0$
\begin{equation}
\Xi = Q_f, \quad \eta=\frac{\pi}{2}
\end{equation}
Hence, the effective ac voltage across the junction reads
\begin{equation}
V_J= Q_f V_{ac}\sin(\omega_0 t)
\end{equation}
The ac-DCB effect alters the Fourier coefficient of the $n$th harmonic of the current by the factor
\begin{equation}\label{fac}
S_n=\frac{Y(n\omega_0)}{Y(n\omega_0)-in\omega_0 C} =\frac{1}{1-n^2-in/Q_f}
\end{equation}
which implies a strong suppression of higher harmonics as shown in Fig.~\ref{fig2}b. This has not been studied experimentally so far. 

Tunneling elements embedded in electromagnetic environments with resonances are currently quite frequently investigated. For these systems the DCB effects predicted here for sinusoidal driving can be significant and requires attention. In this paper we have only addressed explicitly the average current. However, the theory presented has also immediate consequences for the noise spectrum of the current. This and other aspects of the extended DCB theory will be studied in future work.

\begin{acknowledgments}
The author wishes to thank Daniel Esteve, Philippe Joyez, Fabien Portier and Michael Thorwart for inspiring discussions.
\end{acknowledgments}

\end{document}